\documentclass{JHEP3}
\title{Wave function derivation of the JIMWLK equation}
\author{Alexey V. Popov\\Russia, Novgorod State University\\ \email{avp@novgorod.net}}
\abstract{Using the stationary lightcone perturbation theory, we propose the complete and careful derivation the JIMWLK equation. 
          We show that the rigorous treatment requires the knowledge of a boosted wave function with second order accuracy.
          Previous wave function approaches are incomplete and implicitly used the time ordered perturbation theory, which requires a usage of an external target field.
         }
\begin{document}
%\maketitle
\newcommand{\ket}[1]{\ensuremath{|#1\rangle}}
\newcommand{\bra}[1]{\ensuremath{\langle#1|}}
\newcommand{\braket}[2]{\ensuremath{\langle#1|#2\rangle}}
\section*{} 
The JIMWLK evolution equation is the important part of the high energy QCD.
Originally, the equation was derived using the path integral language \cite{JIMWLK,CGC1,Mueller} (see also \cite{CGC05} for generalization).
The clear interpretation of the equation in terms of parton wave function and its evolution was proposed in \cite{Kovner_05}.
The essential advantage of the wave function derivation is the explicit separation between scattering and evolution.
This means that the projectile evolution can be treated irrespective to a target.
Evolution and scattering are distinct and separated parts of the theory. A scattering is Wilson lines 
and an evolution is a Hamiltonian-driven task. Then, in order to calculate amplitude of a realistic process, 
it is necessary to calculate a quasiclassical gauge field generated by projectile and target.

The main aim of this paper is to finish the incomplete derivation which was proposed in \cite{Kovner_05}
where the author used a so called "ingoing" and "outgoing" wave functions. 
Such terms do not exist in the framework of the light cone perturbation theory (LCPT). 
Instead, they are elements of the old-fashion time ordered perturbation theory. 
Similar situation was observed in \cite{Mueller} where different cuts of time ordered diagrams correspond to 
terms of the JIMWLK and BK equations. 
Although the time ordered perturbation theory also can be used for JIMWLK derivation, but this requires 
the explicit usage of an external field, which plays role of time depended potential, and the special technique of
time ordered perturbation theory, which strongly differs with technique that was used in \cite{Kovner_05}.
Summing our remarks, we can say that the work \cite{Kovner_05} contains improper mixing of the two perturbation methods 
which have different areas of applicability.  Neither the time ordered nor stationary perturbation theories were used.
The time ordered perturbation theory can be applied for the evolution of scattering 
amplitude of a projectile in an external field and it is hard to apply it to the evolution of a wave function. 
The wave function evolution contains more information, since it contains information about rapidity of partons. 
So the method of stationary perturbation theory is more general.
In this paper we show that the direct usage of LCPT requires knowledge of the boosted wave function up to second order accuracy of the perturbation theory.
The calculation of the second order correction is not trivial and requires the careful treatment with color indexes of projectile partons.
The usage of pure LCPT allows us to present mathematically complete and rigorous derivation, 
which allows to study only projectile without any mention about a target.
Currently, there is no such derivation in literature. We expect that our derivation is more useful for a first study of JIMWLK.  

We take lightcone coordinates where the coordinate $x^+$ plays role of time in the canonical quantization scheme. 
In order to make all constraints to second class we impose the following common gauge fixing conditions:
\begin{equation} \begin{array}{c}
A^+=0 \\ \partial_\mu A^\mu=\partial_-A^-+\partial_iA^i=0
\end{array} \end{equation} 
So the gauge field $A^\mu$ has only two independent degrees of freedom. After the quantization we have 
\begin{equation}
A^\mu_a(\vec{x},x^-)=\int\limits_{k^+>0}\frac{1}{\sqrt{2k^+}}\left(
         \hat{a}^\dag_{k,a,i}\varepsilon^{*\mu}_i e^{ik^+x^--i\vec{k}\vec{x}}+ 
         \hat{a}_{k,a,i}\varepsilon^{\mu}_i e^{-ik^+x^-+i\vec{k}\vec{x}}
         \right) \frac{d^2k dk^+}{\sqrt{(2\pi)^3}}
\end{equation}
where $i=1,2$ enumerates the gluon polarizations, 
$\varepsilon^\mu_i(k)$ are polarization vectors which obey:
\begin{equation}\begin{array}{c}
\varepsilon^+_i=0 \\
k^+\varepsilon^-_i-\vec{k}\vec{\varepsilon}_i=0\\
\varepsilon^-_i(k)=\frac{1}{k^+}\vec{k}\vec{\varepsilon}_i(k)
\end{array}
\end{equation}
We select real and orthonormal $\vec{\varepsilon}_i$: 
\begin{equation}
  \vec{\varepsilon}_i\vec{\varepsilon}_j=\delta_{ij}
\end{equation}
Useful completeness property is
\begin{equation}
\sum\limits_{i=1,2}(\vec{\varepsilon}_i \vec{u}) (\vec{\varepsilon}_i \vec{v})=\vec{u}\vec{v}
\end{equation}
In the one gluon Hilbert space with the basis $\ket{\vec{k},k^+}$ we shall use basis of transverse coordinates. 
 \begin{equation}
  \ket{\vec{x},k^+}=\int e^{-i\vec{k}\vec{x}}\ket{\vec{k},k^+}\frac{d^2k}{\sqrt{(2\pi)^2}}
 \end{equation}
 \begin{equation}
 \bra{\vec{x},k^+}\vec{y},p^+\rangle=\delta^2(\vec{x}-\vec{y})\delta(k^+-p^+)
 \end{equation}
The gluon creation operator in the given transverse point is
 \begin{equation}
  \hat{a}^\dag_{\vec{x},k^+}=\int e^{-i\vec{k}\vec{x}} \hat{a}^\dag_{k} \frac{d^2k}{\sqrt{(2\pi)^2}}
 \end{equation}
Instead of momentum $k^+$ we shall often use rapidity.
A momentum $p^+$ has rapidity $Y$ relatively $k^+$ if
\begin{equation}
 p^+=k^+e^Y
\end{equation}
The measure is
\begin{equation}
dY=\frac{dk^+}{k^+}
\end{equation} 
If we increase on $\delta Y$ the rapidity of the state $\ket P$, then the wave function will be changed. 
New partons will arise in the new available rapidity window. In spirit of the RG flow we assume that new partons have rapidity much less than primary fast partons. 

Consider a composite dilute projectile $\ket{P}$. We assume that it is moving to the right.
It scatters on a leftmoving dense target state $\ket{T}$. The both two considered states are composed from quarks and glouns.
In the eikonal approximation every fast parton of $\ket{P}$ scatters independently and quasielastic S-matrix is 
\begin{equation} \label{eq_def_S}
 \hat{S}=Pe^{ig\int\limits_{-\infty}^{+\infty}\alpha_a(\vec{x},x^+) \hat{\rho}^a(\vec{x}) dx^+ d^2x }
\end{equation} 
where $\alpha_a(\vec{x},x^+)=A_{T,a}^-(\vec{x},x^+)$ is a classical field generated by target and $\hat\rho^a$ is the color charge density operator.
We wish to know how the scattering amplitude will be changed when we go to the high rapidity
\begin{equation} \label{eq_dS}
  \frac{dS[\alpha]}{dY}\delta Y=\bra {P_{\delta Y}} \hat S \ket {P_{\delta Y}} - \bra P \hat S \ket P 
\end{equation}
where $S[\alpha]=\bra P \hat S \ket P$. The key task is the calculation of state $\ket {P_{\delta Y}}$.
It is necessary to calculate this state with second order accuracy of the QCD perturbation theory.
 \begin{equation}
  \ket {P_{\delta Y}} =\ket {P^{(0)}} + \ket {P^{(1)}} + \ket {P^{(2)}}
 \end{equation}
 \begin{equation} \label{ref7}
 \frac{dS[\alpha]}{dY}=\bra {P^{(1)}} \hat S \ket {P^{(1)}} + \bra {P^{(2)}} \hat S \ket {P^{(0)}} + \bra {P^{(0)}} \hat S \ket {P^{(2)}}
 \end{equation} 
The operator $\hat S$ does not change a parton number and parton coordinates. It acts only on color indexes. 
Hence, it is necessary to know the part of $\ket {P^{(2)}}$ without additional soft gluons.

The lightcone QCD Hamiltonian has complicated structure \cite{Brodsky}. In our calculations there is only one relevant interaction term.
It contains a gluon emission from a color source.
 \begin{equation} \label{eq_1}
  \hat V= -g \int \hat A_\mu^a(\vec x,x^-) \hat J^\mu_a(\vec x,x^-) d^2x dx^-
 \end{equation}
where the current $J^\mu_a$ consists of both quarks and gluons.
A slow gluon has momentum $k^+\ll p^+$, $|\vec k|\ll p^+$ where $p^+$ is the momentum of the source parton.
So only matrix elements like $\bra P J^+_a \ket P$ has nonzero values. The other components $J^\mu_a$ are kinematically suppressed.
The operator $\hat A_-^a(\vec x,x^-)$ creates a soft gluon which has a momentum $k^+\ll p^+$. So in (\ref{eq_1}) we can neglect $x^-$ dependence
in $\hat A_-^a$ and set $x^-=0$.
\begin{equation}
  \hat V= -g \int \hat A_a^-(\vec x,0) \hat \rho^a(\vec x) d^2x 
\end{equation}
\begin{equation}
\rho^a(\vec x)=\int J^+_a(\vec x,x^-) dx^-
\end{equation}
Consider the eigenfunctions of the free Hamiltonian $\ket{\Psi_n^{(0)}}$. Applying the direct analog of the nonrelativistic quantum mechanic perturbation theory, we find 
the first correction
 \begin{equation} \label{eq_2}
 \ket{\Psi_n^{(1)}}= C_{nk}^{(1)}\ket{\Psi_k^{(0)}}
 \end{equation}
 \begin{equation} \label{ref5}
  C_{nk}^{(1)}=\frac{V_{kn}}{E_n^{(0)}-E_k^{(0)}}
 \end{equation}
In our case the momentum $k^-$ plays role of energy. Hence, energy of an emitted slow gluon is much larger than energy of the state $\ket P$ due to 
$k^+\ll p^+$ and the following identity for massless partons
 \begin{equation}
 k^-=\frac{\vec k^2}{2k^+}
 \end{equation}
So the expression (\ref{eq_2}) can be rewritten as
 \begin{equation}
 \ket{P^{(1)}} = -H_0^{-1}V\ket{P^{(0)}}
 =\int \frac{2k^+}{\vec k^2} \frac{\varepsilon^-_i(k) e^{-i\vec k \vec x}}{\sqrt{2k^+}\sqrt{(2\pi)^3}} \hat a^\dag_{\vec k,k^+,i,a} \hat\rho^a(\vec x)\ket{P^{(0)}} d^2\vec k dk^+ d^2x
 \end{equation}  
where the integration on $dk^+$ must be performed on the new opened rapidity window. Going to the coordinate representation, we obtain
 \begin{equation}
 \ket{P^{(1)}} = \int \frac{2}{\sqrt{2k^+}{\sqrt{(2\pi)^5}}} \frac{\vec \varepsilon_i \vec k}{\vec k^2} e^{i\vec k(\vec y-\vec x)} \hat a^\dag_{\vec y,k^+,i,a} \hat\rho^a(\vec x)\ket{P^{(0)}} d^2\vec k dk^+ d^2x d^2y
 \end{equation}  
It is necessary to calculate the integral $\vec I(\vec r)=\int \vec k  k^{-2} e^{i\vec k\vec r}d^2\vec k$.
The value of the integral is a vector so it must has form $\vec r F(r)$, where $F=r^{-2}(\vec r \vec I)$. 
 \begin{equation}
 \vec I(\vec r)=\frac{\vec r}{r^2} \int \frac{\vec k \vec r}{\vec k^2} e^{i\vec k\vec r}d^2k=2\pi i\frac{\vec r}{r^2}
 \end{equation}
where we used the integral $\int\limits_0^\infty e^{it}dt=i$  .
Finally, for the first order we have
 \begin{equation} \label{ref6} 
% \fbox{$
 \ket{P^{(1)}} = \frac{ig}{2\pi\sqrt \pi \sqrt{k^+}}  \int \hat a^\dag_{\vec y,k^+,i,a}  \frac{(\vec y-\vec x) \vec \varepsilon_i }{(\vec y-\vec x)^2} 
 \hat\rho^a(\vec x)\ket{P^{(0)}} dk^+ dx dy %$}
 \end{equation}  
This formula can be easily understood. Each projectile parton with coordinates $\vec x$ can emit a soft gluon into an arbitrary point $\vec y$.
The operator $\hat\rho^a(\vec x)$ measures partons in the projectile. The integration on $x$ and $y$ enumerates all possible emission possibilities.  

In the second order we want to know the state $\ket{P^{(2)}}$ projected on subspace without additional soft gluons.
We can not obtain an analog of the formula (\ref{ref5}) because the energy difference tends to zero. This lack of knowledge is compensated 
by the normalization requirement of a wave function. In our case there is an additional complication. The sector without soft glouns
contains not only state $\ket{P^{(0)}}$ but also state $\ket{P^{(0)}}$ with arbitrary rotations of partons color indexes.
So $\ket{P^{(2)}}$ can differ from $\ket{P^{(0)}}$ on a color rotation.

Let the $\ket{P^{(0)}_n}$ be an orthonormal basis in the space generated by arbitrary transformations of parton color indexes in $\ket{P^{(0)}}$.
Let $\ket{P^{(1)}_n}$ and $\ket{P^{(2)}_n}$ be first and second corrections respectively. 
In order to find \ket{P^{(2)}_n} we require 
 \begin{equation}
 \delta_{mn}= \left(\bra{P^{(0)}_m}+\bra{P^{(1)}_m}+\bra{P^{(2)}_m}\right)\left(\ket{P^{(0)}_n}+\ket{P^{(1)}_n}+\ket{P^{(2)}_n}\right)
 \end{equation}
 which in the second order gives
 \begin{equation}
 \braket{P^{(1)}_m}{P^{(1)}_n}+\braket{P^{(2)}_m}{P^{(0)}_n}+\braket{P^{(0)}_m}{P^{(2)}_n}=0
 \end{equation}
Let $\ket{P^{(2)}_n}=C_{nk}\ket{P^{(0)}_k}$ and $F_{mn}=\braket{P^{(1)}_m}{P^{(1)}_n}$, then we have
 \begin{equation}
   C_{nm}+C_{mn}^*=-F_{mn}
 \end{equation}
It is clear from (\ref{ref6}) that the matrix $F_{mn}$ is hermitian. So the antihermitian part of the matrix $C_{nm}$ can be arbitrary.
But it does not give a contribution to the physical S-matrix. The situation is the consequence of unobservability of an absolute phase 
of a wave function. We assume that the matrix $C_{nm}$ is hermitian. So, we have
 \begin{equation}
 \ket{P^{(2)}_n}=-\frac{1}{2}\ket {P^{(0)}_m} \braket{P^{(1)}_m}{P^{(1)}_n}
 \end{equation}
The basis $\ket {P^{(0)}_m}$ is complete so, using (\ref{ref6}), we have the final expression for the second order correction to the boosted projectile wave function.
 \begin{equation} 
 \ket{P^{(2)}}=-\delta Y\frac{g^2}{2(2\pi)^3}\int\limits_{zxy}^{\phantom{z}}  \frac{(\vec z-\vec y)(\vec z-\vec x)}{(\vec z-\vec y)^2(\vec z-\vec x)^2} 
 \hat \rho^a(\vec y) \hat \rho^a(\vec x)  \ket {P^{(0)}}
 \end{equation}
where $\delta Y$ comes from the integration on the soft gluon momentum $k^+$ in the $\braket{P^{(1)}_m}{P^{(1)}_n}$.
The $\bra {P^{(1)}} \hat S \ket {P^{(1)}}$ term in (\ref{ref7}) can be evaluated with the help of the following identity.
 \begin{equation}
 a_{\vec z,b} \hat S a^\dag_{\vec z,a}= V_{ba}(\vec z) \hat S
 \end{equation}
 where $V_{ba}(\vec z)$ is the one gluon scattering amplitude which is Wilson line in the adjoint representation. 
 \begin{equation}
 V(\vec z) = P e^{ig\int\limits_{-\infty}^{+\infty}\alpha_a(\vec z,z^+)T^a_{AD}dz^+}
 \end{equation}
 and where $T^a_{AD}$ are the generators of the gauge group in the adjoint representation.
 Substituting $\ket {P^{(1)}}$ and $\ket {P^{(2)}}$ into (\ref{ref7}), we obtain the JIMWLK equation
 \begin{equation} \label{eq_dSdY}
 \frac{dS[\alpha]}{dY}=\frac{g^2}{(2\pi)^3}\int\limits_{zxy}^{\phantom{z}} K_{zxy}
 \bra P 
 \left(
 \begin{array}{c}
 - \hat S \hat \rho^a(\vec y)  \hat \rho^a(\vec x) \\
 - \hat \rho^a(\vec y) \hat \rho^a(\vec x) \hat S \\
+2V_{ba}(\vec z) \hat \rho^b(\vec y) \hat S \hat \rho^a(\vec x) 
 \end{array}   \right)
\ket P 
 \end{equation}
\begin{equation}
 K_{zxy}=\frac{(\vec z-\vec y)(\vec z-\vec x)}{(\vec z-\vec y)^2(\vec z-\vec x)^2}
\end{equation}
The equation (\ref{eq_dSdY}) can be expressed in more familiar form. We define the left and right derivative operators $J^a_+(x)$ and $J^a_-(x)$ of color rotations
\begin{equation}
 J^a_\pm(x)=\frac{1}{ig}\frac{\delta}{\delta \alpha_a(x,\pm\infty)} 
\end{equation}
\begin{equation}
\frac{dS }{dY}=
  \frac{g^2}{(2\pi)^3} \int\limits_{zxy}^{\phantom{x}}
  K_{zxy}
  \left[ -J^a_+(x)J^a_+(y)-J^a_-(x)J^a_-(y)+2V_{ba}(z)J^b_+(x)J^a_-(y) \right]   S
\end{equation}

\noindent It should be stressed that it is not necessary to calculate the higher orders of the lightcone perturbation theory with
aim to improve the JIMWLK equation. All higher terms contain higher powers of $\delta Y$ and 
does not give a contribution to the differential equation. The corrections come only from the matrix elements 
of the interaction part of the lightcone QCD Hamiltonian (\ref{eq_1}). From the same reasons we can conclude that an emission 
of two or more gluons into the new opened phase space does not contribute to the evolution equation. In the path integral language
the restriction only to one soft gluon emission corresponds to exactness of the saddle point approximation which used for the integration
of quantum fluctuations in the JIMWLK derivation. One trouble only with the exact evaluation of the saddle point.
Note that all previous speculations are valid only if we study the elastic or quasielastic processes for which the JIMWLK has been initially intended.   
For inelastic inclusive processes we really need to calculate the higher orders perturbation theory terms. 

The lightcone QCD Hamiltonian is plagued by the existence of zero modes and requires a careful gauge fixing procedure. 
One method of studying of all orders was proposed in \cite{Kovner_0501,Kovner_07} and in \cite{CGC05} where path integrals are used.
In the work \cite{alpha_2} was considered the first correction to the JIMWLK equation in the dilute regime. 
In general we can say that the higher corrections gives so called pomeron loops which are very intensively studied last decade.

\bigskip \noindent \textbf{Acknowledgment:} We thank N.V. Prikhod'ko for feedback and useful remarks.


\begin{thebibliography}{99}
\bibitem{JIMWLK} J. Jalilian-Marian, A. Kovner, A. Leonidov and H. Weigert, Nucl. Phys. B504 (1997) 415;
Phys. Rev. D59 (1999) 014014; J. Jalilian-Marian, A. Kovner and H. Weigert, Phys. Rev.
D59 (1999) 014015; H. Weigert, Nucl. Phys. A 703 (2002) 823.
\bibitem{CGC1} E.Iancu, A. Leonidov, L. McLerran, Nucl.Phys. A692 (2001) 583-645, arXiv:hep-ph/0011241.
\bibitem{Mueller} A.H. Mueller, Phys.Lett. B523 (2001) 243-248, arXiv:hep-ph/0110169.
\bibitem{CGC05} Y. Hatta, E. Iancu, L. McLerran, A. Stasto, D.N. Triantafyllopoulos, Nucl.Phys. A764 (2006) 423-459, arXiv:hep-ph/0504182.
\bibitem{Kovner_05} A. Kovner, arXiv:hep-ph/0508232.
\bibitem{Brodsky} S. J. Brodsky, H. C. Pauli, S. S. Pinsky, Phys. Rept. 301, 299 (1998).
\bibitem{Kovner_0501} A. Kovner, M. Lublinsky, Phys.Rev. D71 (2005) 085004, arXiv:hep-ph/0501198.
\bibitem{Kovner_07} A. Kovner, M. Lublinsky, U. Wiedemann, arXiv:0705.1713
\bibitem{alpha_2} J.L. Albacete, N. Armesto, J.G. Milhano, JHEP 0611 (2006) 074, arXiv:hep-ph/0608095.
\end{thebibliography}
\end{document}